%% file: index.tex
\documentclass[pra,aps,twocolumn,nopacs,superscriptaddress,nofootinbib]{revtex4-1}

\usepackage{amsmath}        \usepackage{amssymb}   \usepackage{amsfonts}       \usepackage{bm}       \usepackage{braket}         \usepackage{color}  \usepackage{comment}        \usepackage{dcolumn}  \usepackage{enumerate}      \usepackage{epsfig}  \usepackage{gensymb}        \usepackage{graphicx}  \usepackage{indentfirst}    \usepackage{lmodern}  \usepackage{mathrsfs}       \usepackage{mathtools}  \usepackage{psfrag}         \usepackage{pst-all}   \usepackage{soul}           \usepackage{xcolor}
\usepackage{upgreek}        \usepackage{bbm}       
\usepackage{float}          \usepackage{lipsum}

\usepackage[colorlinks,linkcolor=blue,citecolor=blue,urlcolor=blue,hyperindex,driverfallback=dvipdfm]{hyperref}  \usepackage[T1]{fontenc}

    \newcommand{\im}{\text{i}}
    \newcommand{\vt}[1]{\textbf{#1}}

    \usepackage{afterpage}

\makeatletter
\def\widetext@line{} 
\def\endwidetext@line{} 
\makeatother 

\begin{document}

\def\bibsection{\section*{\refname}}

\title{Thermal diffuse scattering in TEM: complex absorptive potentials compared to the frozen phonon model}

    \author{Martin Hájek} 
        \affiliation{Institute of Physical Engineering, Brno University of Technology, Technická 2, 616 69 Brno, Czech Republic}
        
    \author{Ján Rusz}
        \affiliation{Department of Physics and Astronomy, Uppsala University, Box 530, 75121 Uppsala, Sweden}

    \begin{abstract}
        \input{Abstract}
    \end{abstract}
    
    \maketitle
    \date{\today}
      
    \input{Introduction}

    \input{Theoretical_framework}

\input{Results_and_discussion}

    \input{Conclusion}
    
    \bibliographystyle{apsrev}
    \bibliography{Bibliography.bib} 

\newpage
 
\end{document}

%% file: Abstract.tex
In transmission electron microscopy, electrons undergo inelastic scattering primarily through phonon excitations, known as thermal diffuse scattering. To capture the inelastic scattering effects on the elastic scattering component, absorptive effects must be included in the modeling of electron propagation, accounting for the gradual depletion of the elastic channel of the electron beam. Several approaches to modeling this absorption exist. In this paper, we compare the widely used complex absorptive potentials method to the more elaborate frozen phonon model, based on correlated atomic motion and on the Einstein model of atomic motion.

%% file: Introduction.tex
\section{Introduction}
Inelastic scattering on lattice vibrations, also known as thermal diffuse scattering (TDS), plays a crucial role in high-resolution imaging and electron diffraction \cite{Reimer, HRTEM, Krivanek_phonon}. TDS shows up as a diffuse background in diffraction patterns, and in simulations, the TDS background may differ in shape, intensity, and characteristic features (such as Kikuchi lines) depending on the model used \cite{chuvilinCBED2005, mendis_phase_2023}. Electrons that are inelastically scattered to high scattering angles make up a significant portion of the intensity signal in the dark-field \cite{mullerTDS}, especially the high-angle annular dark-field (HAADF) \cite{lugg2015}. Thus, a rigorous approach for inelastic scattering implementation is needed, not only to simulate intensity distribution correctly, but the electron imaging and high-resolution electron imaging in general \cite{subangstrom}.

The complex absorptive potential (CAP) method, first described by Yoshioka \cite{Yoshioka}, views inelastic scattering as the excitation of the crystal and uses many-body quantum mechanics to describe absorption effects in the elastic channel caused by the inelastic electron scattering. In this framework, TDS distribution can not be directly mapped, but its intensity contribution is effectively removed from the purely elastic part of the total intensity. Yoshioka's theory has later been extended by Martin et al.\ \cite{martin2009} and Forbes et al.\ \cite{QEP, forbes2016}. Martin et al.\ present and compare the complex absorptive potential model for correlated atomic motion and independent harmonic atomic motion, also known as the Einstein model, and claim that both approaches are in agreement provided a large detector collection area. The large integration area, experimentally manifested as a large detector, furthermore shows up as an argument for the valid use of a local approximation of the otherwise non-local character of the complex absorptive potential in the work of Findlay et al.\ \cite{findlay_stem}. Practical implementations of the CAP can be performed with parametrization of the absorptive form factors presented in the work of Weickenmeier and Kohl \cite{weickenmeier1991} or Peng et al.\ \cite{peng1996}.

Forbes et al.\ present a model for the quantum excitation of phonons for multiple thermal scattering events in the Born-Oppenheimer approximation. Furthermore, in the CAP model, Hall et al.\ \cite{hallTDS} state that the effects of TDS contribute non-negligibly to the absorption and strongly depend on the atomic number of the element in the crystal. Forbes et al., as well as Van Dyck \cite{vandyck}, show that the results of the quantum mechanical model agree well with the results of the frozen phonon model. 

The frozen phonon model (FPM) is a semi-classical approximative model that incorporates electron propagation through a static lattice \cite{loane_thermal_1991}. Since the velocity of the electron is significantly higher relative to atomic motion in the lattice, for the traversing electron, the lattice appears as if it was ``frozen in time''. The so-called multislice calculations then trace the evolution of the electron wavefunction traversing through the sample by dividing it into a finite number of slices along the z-axis and iteratively scattering on the projected potential and propagating the wave slice by slice through the whole sample, resulting in a final wave function in the form of a diffraction pattern \cite{Kirkland}.

The diffraction pattern obtained via the FPM then arises from incoherent averages of the propagated wavefunctions over multiple static lattices, so-called ``snapshots'' \cite{paul2020}. This approach originates in the quantum excitation of phonons (QEP) theory by Forbes et al.\ \cite{QEP}. Furthermore, unlike for the complex absorptive model, the TDS distribution is obtainable in the frozen-phonon model framework \cite{martin2009}. The snapshots represent the crystal configuration of atoms at different points in time. There is a sufficient time gap between each snapshot, which corresponds to experimental reality, as each traversing electron ``sees'' a different crystal configuration. The practical difference in the FPM simulation then arises from the method choice in the snapshot generation. One can utilize interatomic potentials in a molecular dynamics (MD) simulation to create correlated atomic motion in the sample or use a simplified model of harmonic atomic motion, the Einstein model. 

Although the CAP model is more computationally efficient than FPM, since it does not require MD or multiple snapshots, one must be cautious in employing it and ensure that the conditions of validity are met \cite{Mendis2019_plasmonMS}. We present a qualitative comparison between the widely used CAP model and FPM on diffraction patterns of a diamond and STO crystals under parallel illumination conditions and investigate the factors that affect the agreement of the models, as well as the conditions under which these models do or do not agree. We mainly pay attention to differences in the elastic channel intensities, particularly the Bragg-spot intensities, as Bragg-intensity discrepancies are most influential in the image formation. For well-established interatomic potentials and subsequently the MD simulations, the correlated atomic motion model should, in theory, be the best in describing the reality of atomic motion \cite{vandyck, loane_thermal_1991, chen_comparison_2023}. We thus consider the correlated atomic motion FPM to be the most precise of the three models used.

%% file: Theoretical_framework.tex
\section{Theoretical framework and methods}

\subsection{Underlying equations}

In derivation presented in Ref.~\cite{findlay_stem}, the otherwise non-local absorptive potential $W(\vt{R},\vt{R}^\prime)$ is conveniently approximated to a local absorptive potential $V^\prime(\vt{R})$ (provided large detector as stated) representing the inelastic losses in the total intensity in plane $\vt{R}=(x, y)$: 
\begin{equation}
    W(\vt{R},\vt{R}^\prime)\approx 2 V^\prime(\vt{R})\delta(\vt{R}-\vt{R}^\prime).
\end{equation}
In Ref.~\cite{weickenmeier1991}, the complex potential is obtained through parametrized absorptive form factors; the imaginary part of the crystal potential then depletes the elastic wavefunction as it propagates. This approach was utilized in the \textsc{DrProbe} \cite{barthel_drprobe} multislice simulation solution (used in our workflow), where the slices are discrete representations of the projected potential:
\begin{equation}
    V_{\text{proj}}(\vt{R})=V(\vt{R})+\im V^\prime(\vt{R}).
\end{equation}
The wave function evolution traced through the crystal is governed by the multislice solution introduced in \cite{multislice1} and further discussed in \cite{Kirkland, dynamicaldiffraction, cai_computational_2009}. The propagation from slice $j$ to $j+1$ follows:
\begin{equation}\label{MSA}
    \psi_{j+1}(\vt{R})=FT^{-1}\left[P_j(\vt{k})FT\left(T_j(\vt{R})\psi_j(\vt{R})\right)\right],
\end{equation}
where $T(\vt{R})=\exp\left(\im \sigma V_{\text{proj}}(\vt{R})\right)$ is the transmission function responsible for the phase shift and wave function attenuation in the case of the imaginary potential component. $\sigma=me\lambda/(2\pi\hbar^2)$ is the interaction constant, $m$ relativistic electron mass, $\lambda$ electron wavelength and $P_j$ is Fresnel propagator function.

In frozen phonon models, the imaginary part in the projected potential is absent, but the multislice approach remains qualitatively the same. In the semi‑classical FPM, each multislice calculation through a frozen snapshot is purely elastic, and absorption arises only after averaging. Under the QEP of Forbes et al. \cite{QEP}, the elastic channel of the FPM is the coherent (amplitude‑level) average over $N$ snapshots \cite{paul2023, niermann_2019}:
\begin{equation}
    I_{\text{elastic}}(\vt{R})=\left|\frac{1}{N}\sum_{n=1}^N\psi_n(\vt{R})\right|^2,
\end{equation}
while the total intensity is the incoherent average
\begin{equation}
    I_{\text{total}}(\vt{R}) = \frac{1}{N}\sum_{n=1}^N\vert\psi_n(\vt{R})\vert^2.
\end{equation}
and the TDS (inelastic) component is the difference
\begin{equation}
    I_{\text{TDS}}=I_{\text{total}}-I_{\text{elastic}}.
\end{equation}

\subsection{Simulation processes}

Snapshots used in the FPM are of two kinds. The first set of snapshots is of correlated atomic motion obtained via molecular dynamics LAMMPS \cite{LAMMPS} simulation, and the second set of snapshots corresponds to the Einstein model, i.e., independent harmonic motion of atoms. Note that, both the FPM in Einstein model and the CAP model require the knowledge of Debye-Waller factors \cite{peng1996}. The resulting diffraction patterns are subsequently obtained via multislice calculations in the software \textsc{DrProbe} \cite{barthel_drprobe} on the crystal snapshots. 

We performed a series of multislice calculations regarding the correlated atomic motion FPM (will also be referred to as MD-FPM), additionally, FPM with Einstein distribution and complex absorptive model, which is represented in the calculation by parametrized complex form factors \cite{weickenmeier1991}. We obtained the intensity distribution in diffraction patterns, primarily the elastic component of the intensity, which is depleted due to the losses into the inelastic channel. The definition of the crystal superlattices and MD simulation took place in the LAMMPS software \cite{LAMMPS}. One corresponding static lattice configuration at equilibrium was retained for use in the CAP model.

The comparison of the models is presented on crystal samples of diamond (C) and subsequently, strontium titanate (STO). These two materials are chosen deliberately, since diamond serves as a light-element reference (C, atomic mass 12.011 u) in which TDS absorption is modest compared to STO, which contains heavy elements, namely Sr (87.62 u), Ti (47.867 u), and the lighter O (15.999 u). Furthermore, we extended the analysis to the case of STO, because the STO anisotropic atomic vibrations represent an additional variable absent from the standard isotropic treatment employed in both the CAP and the Einstein model, allowing us to investigate the limit cases of these approximations for structurally complicated materials \cite{yang_phonon_2024, deepmd}. Lastly, the effect of variable material thickness on the intensity depletion — among all models — is discussed.

The diamond superlattice dimensions were set to 3.5\,nm$~\times~$3.5\,nm$~ \times~$50\,nm, (10$\times$10$\times$140 unit cells) and the lattice parameter of the face-centered cubic unit cell is set to 0.3572\,nm, which is a lattice parameter value resulting from the slight change in crystal volume via a separate MD simulation with $NPT$ thermostat and the lattice parameter differs slightly from the widely referenced value for diamond lattice parameter 0.3567\,nm at room temperature \cite{diamond_lattice}. We then proceeded with an $NVT$ thermostat MD simulation consisting of initial thermal relaxation and the main simulation, which yields the snapshots. The thermal relaxation is responsible for the removal of the initial internal stress of the crystal at the simulation temperature and the delocalization of atoms around their respective equilibrium positions.
The initial process consisted of letting the MD simulation run for a few thousand timesteps (one timestep was set to 1 fs) from the starting point, which was the lattice in equilibrium. In this way, we ensured that the crystal had more than enough time for sufficient thermal relaxation \cite{paul2020, paul2021}. This crystal state then acts as a starting point for the main data-yielding simulation.

The supercell is oriented in such a way that the (001) surface is perpendicular to the subsequent electron propagation direction, the $z$-axis. The interatomic potential for the MD simulation in diamond was the carbon (C-C) Tersoff potential \cite{tersoff_empirical_1988, tersoff_new_1988}. The crystal was kept at 300\,K via the $NVT$ thermostat \cite{thermostat}, embedded in the LAMMPS code, and during the data-yielding simulation, the lattice snapshots were read out every 1000 timesteps.

We define the crystal of STO with a cubic unit cell lattice parameter 0.3905\,nm, and use the MD calculations for the STO-specific machine-learned deepMD interatomic potential \cite{deepmd}. The calculations with deepMD are computationally more complex, which leads to a significant increase in calculation time. To account for this, the size of the supercell needs to be sized reasonably. The supercell dimensions are 3.1\,nm$~\times$~3.1\,nm$~\times~$30\,nm, (8$\times$8$\times$80 unit cells), with the supercell (001) surface aligned with the $z$-axis. The MD simulation process was analogous to the one presented in diamond, at a temperature of 300\,K, with sufficient thermal relaxation of the crystal.

A total of 100 snapshots were used in the multislice simulations of correlated atomic motion FPM, both for diamond and STO specimens, and 50 snapshots in the case of the Einstein FPM. As will be seen below, the Einstein FPM numerically converges faster than MD-FPM, and thus, this lower number of snapshots is sufficient. 

Multislice algorithm simulations were performed to model electron wave propagation through the crystal specimens under high-resolution transmission electron microscopy conditions. The projected electrostatic potential of each specimen was discretized on a 640 $\times$ 640 pixel grid to ensure adequate real-space sampling. For the diamond structure, the specimen thickness was divided into 1120 slices along the beam propagation direction, whereas 1280 slices were used for the STO structure to account for the individual unit cells and the thickness of both specimens. Electron scattering was simulated using an aberration-free electron wave function at an accelerating voltage of 300 kV under parallel illumination conditions, thereby isolating the effects of thermal atomic motion and structural correlations on the resulting exit wave and image formation \cite{allen_inelastic, allen_modelling, van_den_broek_fdes_2015}.

One of the key points to determine the discrepancies between the models is to ensure the multislice simulation data convergence in the FPM arising from incoherent averages over multiple snapshots. In theory, for an infinite number of snapshots, the elastic channel should contain the Bragg spots only. Thus, by adding more snapshots to the simulation, the residual TDS noise in the elastic channel gradually disappears. We also analyze the five most intense Bragg spots and their convergence as we add snapshots, and determine the standard deviation of the Bragg spot intensities over the last 50 snapshot additions. The standard deviation of the central Bragg spot for diamond is $5.65\cdot10^{-5}\, I_0$ and $1.65\cdot10^{-5}\, I_0$ for MD-FPM and Einstein FPM, respectively. For STO, the values are: $6.73\cdot10^{-5}\, I_0$ for MD-FPM and $2.71\cdot10^{-5}\, I_0$ for Einstein FPM. The standard deviation will serve as a threshold to which we compare the differences between various approaches. Differences that fall below the threshold should be viewed as inconclusive \cite{biskupek_evaluation_2007}.

%% file: Results_and_discussion.tex
    \begin{figure*}[t]
        \centering
        \includegraphics[width=\textwidth]{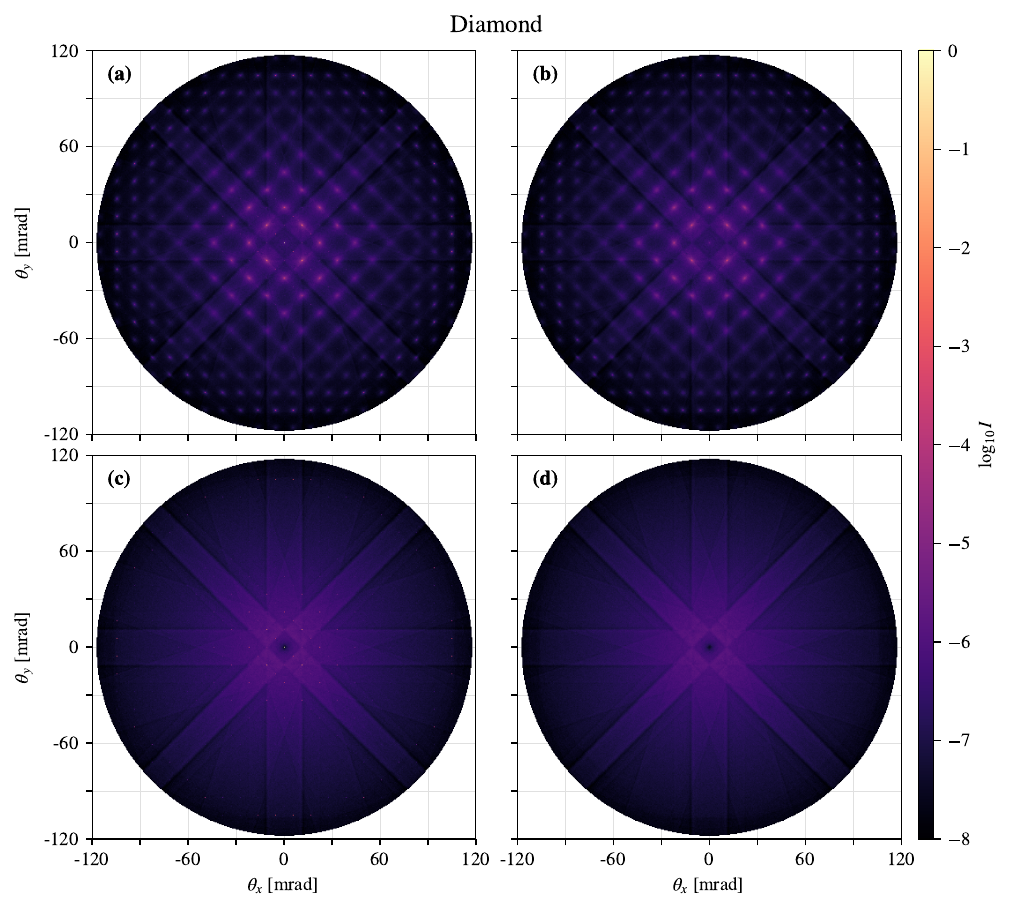}
        \caption{Two-dimensional total-intensity and TDS diffraction-pattern intensity for diamond at 50\,nm specimen thickness and at 300\,kV under parallel illumination.  The first column displays the total intensity ($I_\text{tot}$) and the second column displays the TDS component. (a) MD-FPM total intensity, (b) MD-FPM TDS, (c) Ein-FPM total intensity, (d) Ein-FPM TDS. Intensities are plotted on a decimal logarithmic colour scale ($\log_{10} I$, with $I$ normalised to $I_0$).}
        \label{fig1Diamond}
    \end{figure*}
    
    \begin{figure*}[]
        \centering
        \includegraphics[width=\textwidth]{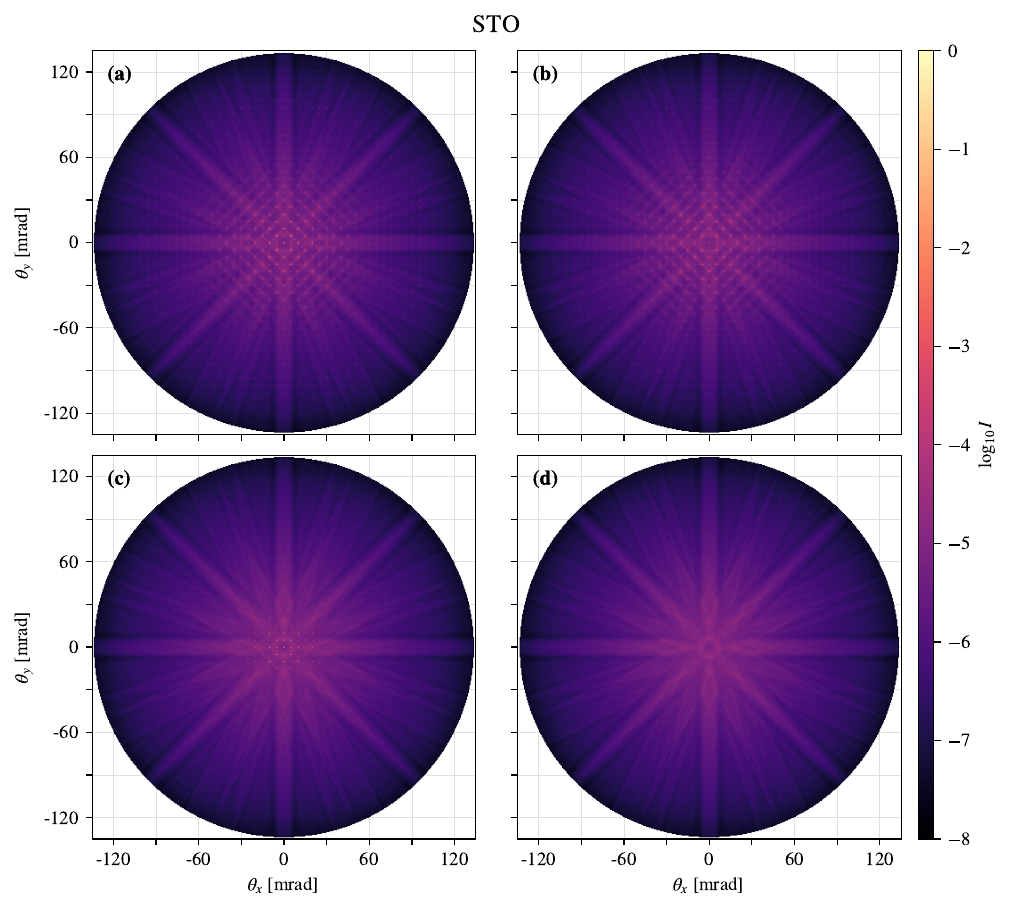}
        \caption{Two-dimensional total-intensity and TDS diffraction-pattern intensity maps for strontium titanate (STO) at 30\,nm specimen thickness, 300\,kV under parallel illumination.  The first column displays the total intensity ($I_\text{tot}$) and the second column displays the TDS component: (a) MD-FPM total intensity, (b) MD-FPM TDS, (c) Ein-FPM total intensity, (d) Ein-FPM TDS. Intensities are plotted on a decimal logarithmic colour scale ($\log_{10} I$, with $I$ normalised to $I_0$).}
        \label{fig2STO}
    \end{figure*}

\section{Results and discussion}

In section \ref{sA}, we show and discuss the two-dimensional total-intensity and TDS diffraction patterns for diamond and STO, respectively, obtained from the two frozen phonon models. The diffraction patterns are then further analyzed in accordance with the large detector theory in section \ref{sB} by convolution of the patterns from the elastic channel annular intensity profiles obtained by integration of the diffraction pattern over rings of constant scattering angle $\theta$ as the integration variable, and further, the elastic-channel pairwise model comparisons are presented. The thickness dependence of these profiles for all models is discussed in section \ref{sC}.

    \begin{figure*}[]
        \centering
        \includegraphics[width=\textwidth]{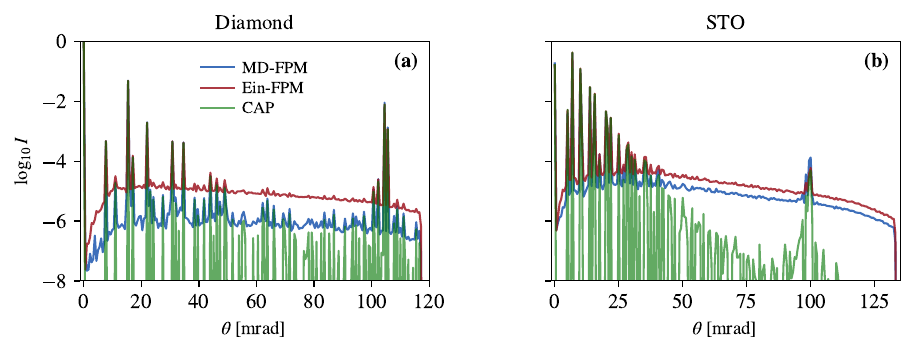}
        \caption{Annularly integrated elastic-channel intensity profiles ($I(\theta)$) at maximum specimen thickness, for all three absorption models. Intensities are obtained by azimuthal integration of the elastic channel diffraction pattern over rings of constant scattering angle ($\theta$) and are plotted on a decimal logarithmic scale, normalised to the incident intensity ($I_0$). (a) Diamond (50\,nm) and (b) STO (30\,nm). Curves: MD-FPM (blue), Einstein FPM (red), and CAP (green). Sharp peaks correspond to Bragg reflections of the (001)-oriented supercell and to higher-order Laue zone (HOLZ) reflections near ($\theta \approx 100\,\text{mrad}-110\,\text{mrad}$; the broad baseline seen in the MD-FPM and Ein-FPM curves is the residual TDS noise (caused by the finite coherent averaging), which is absent from the CAP profile.}
        \label{fig3Profiles}

    \end{figure*}

\subsection{Total intensity and TDS distribution}\label{sA}

Figures~\ref{fig1Diamond} and~\ref{fig2STO} show the total intensity — corresponding to the quantity recorded on the detector — and the TDS component obtained, according to the QEP model \cite{QEP}, by subtracting the elastic channel from the total intensity, for the MD-FPM and the Einstein FPM applied to diamond and STO, respectively. In both materials, the total intensity is dominated by the central beam, and a logarithmic intensity scale is used throughout to reveal the structure of the diffuse background. The CAP model is not represented in Figs.~\ref{fig1Diamond} and~\ref{fig2STO}, as the TDS spatial distribution is not accessible within the CAP framework; its elastic channel — containing the Bragg spots only — is instead examined below in Figs.~\ref{fig3Profiles}~and~\ref{fig4Profiles_diff}.

For 50\,nm thick diamond Fig.~\ref{fig1Diamond}, the total intensity patterns of both FPM variants exhibit characteristic Kikuchi lines arising from the Bragg re-scattering of diffusely scattered electrons \cite{Reimer, hallTDS}. Higher-order Laue zone (HOLZ) reflections are visible near $\theta\approx100\,\text{mrad}$ \cite{Reimer, chuvilinCBED2005}. The TDS distributions from the two models are qualitatively similar, yet the intensity in the immediate vicinity of the Bragg spots is visibly more spread out in the correlated atomic motion model relative to the Einstein model, a direct consequence of the interatomic phonon correlations that the MD snapshots capture and that the independent harmonic oscillator Einstein treatment does not \cite{kong_phonon, loane_thermal_1991}.

In case of STO results, displayed in Fig.~\ref{fig2STO}, the qualitative character of the diffraction patterns is analogous, but there are systematic differences between the two materials. Most notably, despite the STO supercell being only 30\,nm thick compared to 50\,nm for diamond, a substantially larger fraction of the total intensity is redistributed into non-central Bragg spots and the TDS background, which may also be attributed to the stronger scattering induced by heavy elements present in the STO crystal \cite{hallTDS, bethemethod}.

\subsection{Elastic channel intensity profiles: model comparison}\label{sB}

The annularly integrated elastic-channel profiles from Fig.~\ref{fig3Profiles} are subsequently used to show model differences via mutual subtraction among all three models in Fig.~\ref{fig4Profiles_diff}, where the absolute differences are displayed as red solid lines on a logarithmic scale; the relative differences, evaluated at the relevant positions of Bragg reflections, are indicated by blue cross markers. Note that all comparisons are to be evaluated against the thresholds established from the Bragg-spot convergence standard deviations of MD-FPM and Einstein-FPM (marked orange), and differences below the threshold level should be viewed as statistically less significant or inconclusive.

For diamond, the three elastic intensity profiles are in broad qualitative agreement, yet some discrepancies appear clearly above the noise floor. The largest differences arise between the MD-FPM and the other two models. At the central Bragg spot ($\theta=0\,\text{mrad}$), the absolute discrepancy between FPM-MD and the two other models reaches approximately $10^{-3}I_0$. Key finding is that this difference does not disappear in high scattering angles: at the HOLZ positions near $\theta \approx 110\, \text{mrad}$ the absolute error remains comparable in value to those at the central beam or at low angles. Because HOLZ reflections carry far less absolute intensity than the zero-order Bragg spots, their relative error is disproportionately large.

The Einstein FPM and the CAP show considerably better mutual agreement than either of them does with the MD-FPM, consistent with the theoretical expectation that both models converge to equivalent results for a sufficiently large detector collection area. To examine this quantitatively, the diffraction patterns were convolved with circular apertures of diameters 27, 55, and 110\,mrad, depicted in Fig.~\ref{fig5convolution}, representing progressively larger effective detector areas. For the 110\,mrad kernel, the maximum discrepancy among all three models is reduced to approximately $10^{-4.5}I_0$, demonstrating that a large collection area substantially suppresses the inter-model disagreement in accordance with the prediction of Martin et al.\ \cite{martin2009}.

In STO, the inter-model discrepancies are appreciably larger than those found for diamond at equivalent scattering angles. At low scattering angles, the absolute differences between models reach values in the range $10^{-1.5}I_0$ -- $10^{-2}I_0$, corresponding to errors of the order of units of percent. At high scattering angles in the HAADF regime, the absolute difference as well as the relative difference between the correlated FPM and the CAP remain significant.
This finding, however, may be overshadowed by the fact that some of the model absolute difference values fall under the previously established error validity threshold. 

Two distinct physical mechanisms drive the enhanced disagreement of the models. First, the strong atomic-number dependence of TDS absorption established by Hall and Hirsch \cite{hallTDS} means that the mismatch between the absorption treatment in the CAP and the full scattering physics captured by the FPM is amplified for heavy elements. Second, the anisotropic atomic vibrations of the STO perovskite structure are not described by the isotropic Debye-Waller factors employed in the CAP or the Einstein FPM; the correlated MD snapshots capture this anisotropy naturally, whereas the two simpler models cannot, introducing a systematic discrepancy that is absent in the case of diamond. Taken together, these factors make the STO results a sensitive indicator of the conditions under which the CAP model ceases to be a reliable approximation.

\subsection{Thickness dependence}\label{sC}
Figure~\ref{fig6thickness} presents the elastic-channel annular intensity profiles as two-dimensional maps of scattering angle versus specimen thickness, which are easily accessible via multiple wave function readouts during the multislice simulation process. Fig.~\ref{fig7thicknessDiff} shows the corresponding pairwise differences between these maps, evaluated across 20 sampled thickness values for both materials and all three models. For both diamond and STO, all three models reproduce the characteristic oscillatory redistribution of Bragg-spot intensity with increasing thickness, arising from the dynamical redistribution of intensity among reflections via multiple scattering. In the FPM results, a monotonic increase in the diffuse TDS background level with thickness is apparent, consistent with the cumulative increase of multiple inelastic scattering probability along the propagation path; this background is inherently absent from the CAP maps.

For diamond, the inter-model difference maps in Fig.~\ref{fig7thicknessDiff}(a--c) are distributed relatively uniformly as a function of thickness without a strong systematic trend, reflecting the weak TDS absorption of the light carbon lattice. In STO (panels d--f), the difference maps reveal a gradual increase in the discrepancy between the MD-FPM and the other two models at low to mid scattering angles ($\theta \approx 50\,\text{mrad}$) and at high scattering angles near $\theta \approx 100\,\text{mrad}$ as specimen thickness grows, a notable increase in discrepancies is observed in the HOLZ Bragg positions. This progressive divergence indicates that the models accumulate distinct elastic-channel phase information during propagation through the STO lattice, and that the neglect of interatomic phonon correlations and vibrational anisotropy becomes increasingly consequential for thicker specimens. These results highlight that the reliability of the CAP model indicates sensitivity not only to atomic number but also to specimen thickness — a practically important consideration when interpreting elastic scattering of thick samples containing heavy elements.

\begin{widetext}

    \begin{figure}[H]
            \centering
            \includegraphics[width=\textwidth]{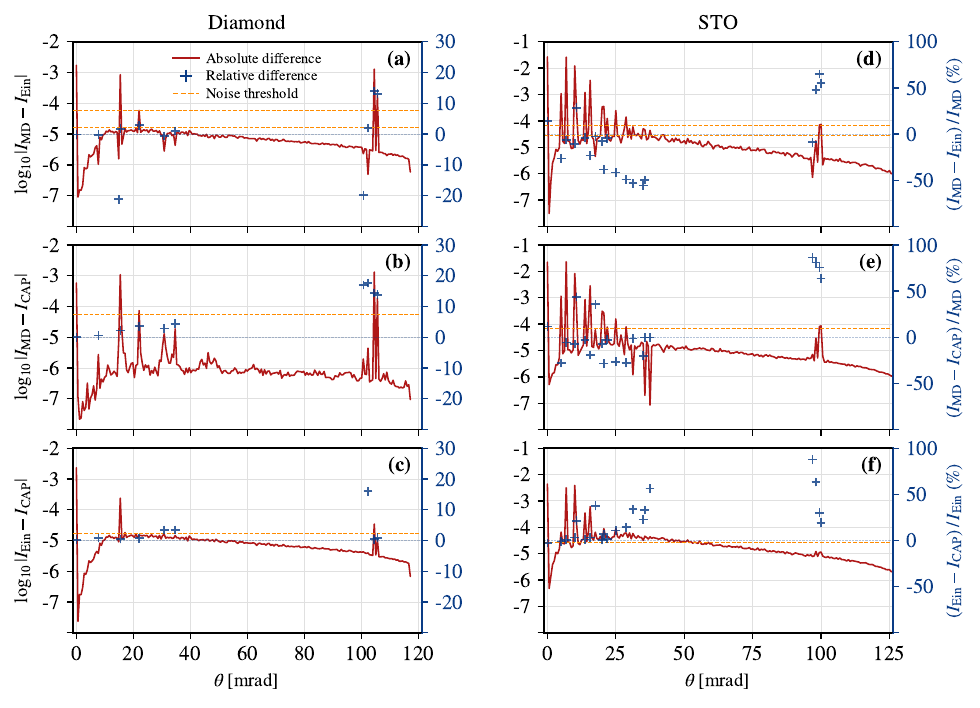}
            \caption{Pairwise inter-model differences of the elastic-channel annular profiles of Fig.~\ref{fig3Profiles}. (a)–(c) Diamond at 50\,nm; (d)–(f) STO at 30\,nm. Red solid curves: absolute difference ($|I_A-I_B|/I_0$) on a decimal logarithmic scale (left ordinate, ($\log_{10}|I_A-I_B|$)). Blue "+" markers: relative difference ($(I_A-I_B)/I_A)$ expressed in per cent (right ordinate), evaluated at the positions of the principal Bragg reflections. Panels display: (a) (A,B=MD-FPM, Ein-FPM); (b) MD-FPM, CAP; (c) Ein-FPM, CAP — for diamond. (d) MD-FPM, Ein-FPM; (e) MD-FPM, CAP; (f) Ein-FPM, CAP — for STO. The horizontal orange dashed lines mark the statistical convergence thresholds obtained from the standard deviation of the cumulative mean of the central Bragg-spot intensity over the last 50 snapshot additions; where two lines are shown, the upper corresponds to the MD-FPM threshold and the lower to the Einstein-FPM threshold.} 
            \label{fig4Profiles_diff}
    \end{figure}

    \begin{figure*}[]
        \centering
        \includegraphics[width=0.95\textwidth]{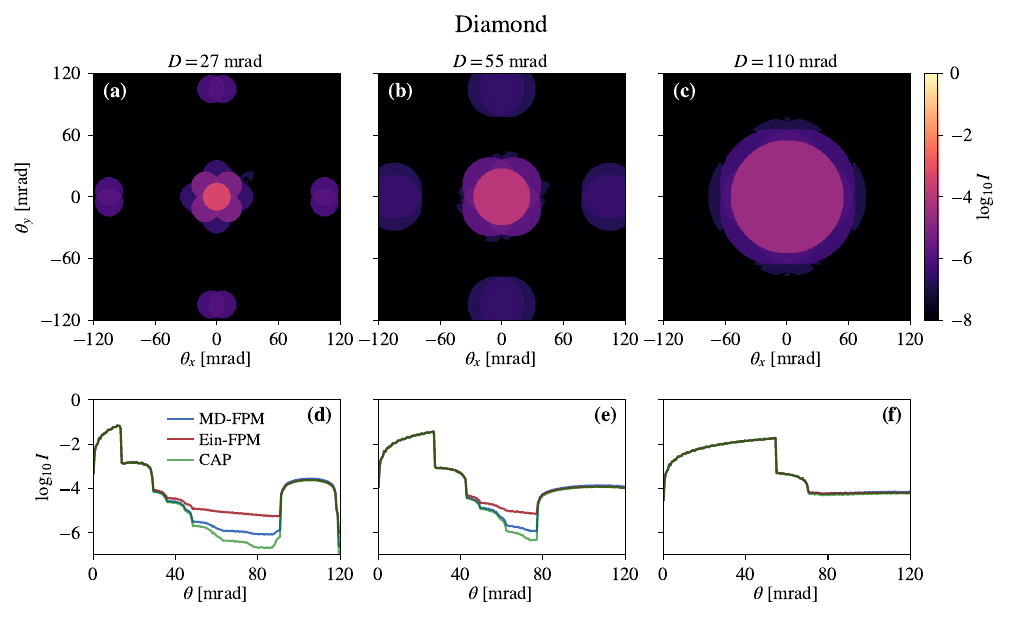}
        \includegraphics[width=0.95\textwidth]{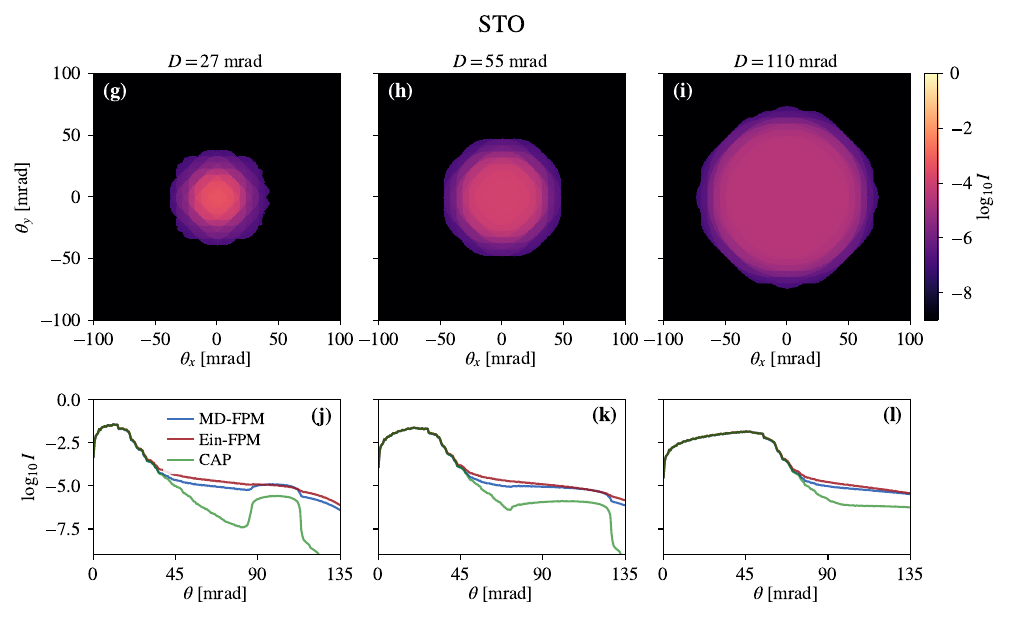}
        \caption{Elastic diffraction patterns of the MD-FPM convolved with circular apertures of diameter $D = 27$, 55, and 110\,mrad (panels a–c), and corresponding annular-integrated elastic intensity profiles of all three models after convolution with the respective aperture (panels d–f) for diamond. The progressive increase of the effective detector collection area reveals increasing agreement between the FPM models and the CAP model. Diamond (C), (001) orientation, 300\,kV, specimen thickness 50\,nm. Analogously, panels g-i show convolved elastic diffraction patterns and j-l the profiles of the MD-FPM for the STO crystal in (001) orientation, 300\,kV, specimen thickness 30\,nm.}
        \label{fig5convolution}

    \end{figure*}

    \begin{figure}[H]
            \centering
            \includegraphics[width=0.92\textwidth]{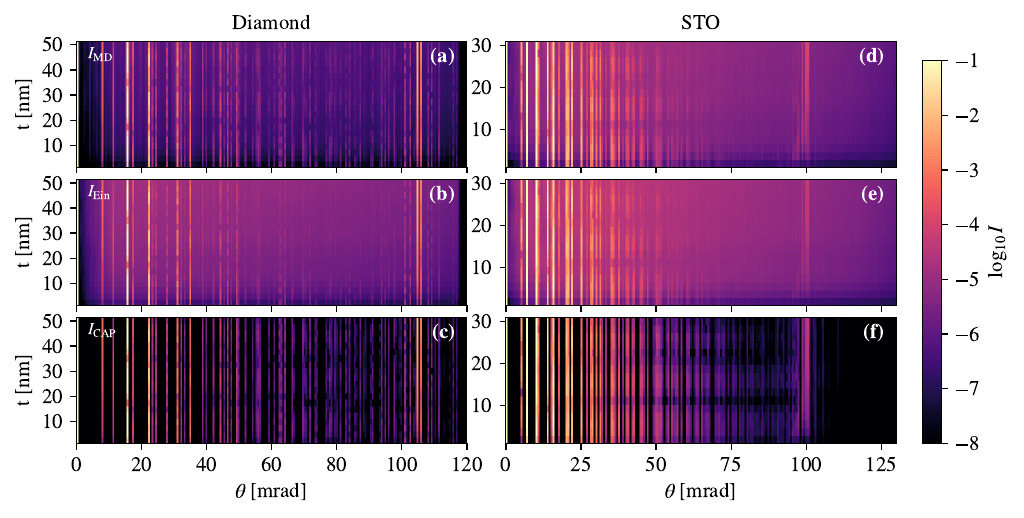}
            \caption{Two-dimensional maps of the annular elastic-channel intensity ($I(\theta,t)$) as a function of scattering angle ($\theta$) and specimen thickness ($t$), for all three models and both materials. Maps are constructed from 20 equispaced wavefunction readouts during multislice propagation. (a)–(c) Diamond (($t$) up to 50\,nm): (a) MD-FPM, (b) Einstein FPM, (c) CAP. (d)–(f) STO ((t) up to 30\,nm): (d) MD-FPM, (e) Einstein FPM, (f) CAP. Intensity is plotted on a common decimal logarithmic colour scale ($\log_{10} I$), normalized to ($I_0$). Vertical stripes identify Bragg reflections whose intensity oscillates with thickness due to dynamical (Pendellösung-like) redistribution; the gradual brightening of the inter-Bragg background with increasing ($t$) in panels (a), (b), (d) and (e) reflects the cumulative build-up of the residual TDS level in the FPM simulations — a feature inherently absent from the CAP maps (c) and (f).}
            \label{fig6thickness}

    \end{figure}
 
    \begin{figure}[H]
            \centering
            \includegraphics[width=0.92\textwidth]{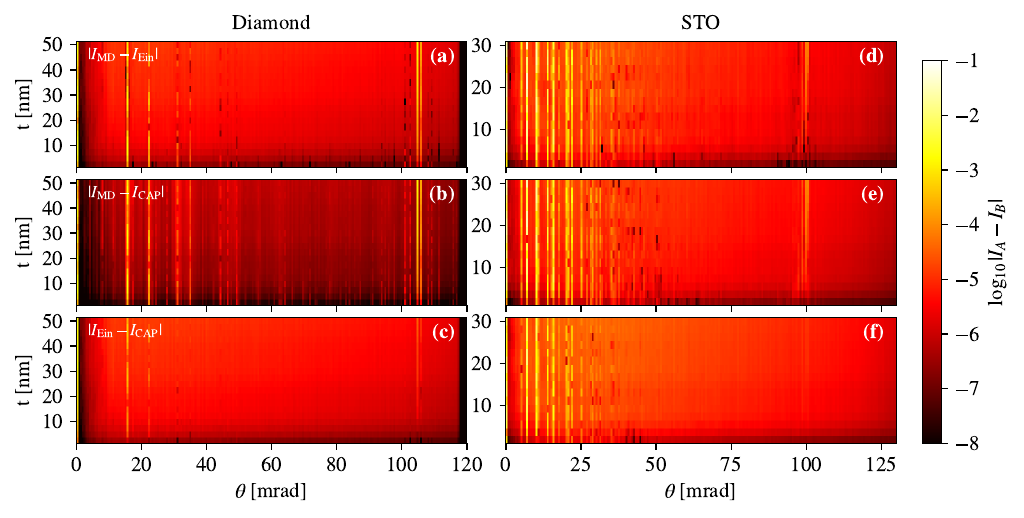}
            \caption{Two-dimensional maps of the pairwise absolute differences ($|I_A(\theta,t)-I_B(\theta,t)|$) between the elastic-channel annular intensity profiles. (a)–(c) Diamond; (d)–(f) STO. The panels display: (a) ($|I_\mathrm{MD}-I_\mathrm{Ein}|$), (b) ($|I_\mathrm{MD}-I_\mathrm{CAP}|$), (c) ($|I_\mathrm{Ein}-I_\mathrm{CAP}|$) for diamond, and (d) ($|I_\mathrm{MD}-I_\mathrm{Ein}|$), (e) ($|I_\mathrm{MD}-I_\mathrm{CAP}|$), (f) ($|I_\mathrm{Ein}-I_\mathrm{CAP}|$) for STO. Differences are plotted on a common decimal logarithmic color scale ($\log_{10}|I_A-I_B|$), normalized to ($I_0$). The color encodes the magnitude of the model disagreement at each ($\theta,t$) coordinate.}
            \label{fig7thicknessDiff}

    \end{figure}
    
\end{widetext}

%% file: Conclusion.tex
\section{conclusions}

We have presented a systematic comparison of three absorption models for the description of elastic scattering in high-energy electron diffraction simulations: the complex absorptive potential model (CAP), the frozen phonon model with correlated atomic motion from molecular dynamics simulations (MD-FPM), and the frozen phonon model with independent harmonic atomic oscillations following the Einstein approximation (Einstein FPM). Multislice calculations at 300 kV for diamond and STO — a light-element cubic crystal and a heavy-element crystal with anisotropic atomic vibrations — were chosen to represent opposite extremes of the conditions under which the models are expected to agree or disagree.

The key finding is that non-negligible differences exist between the CAP and the frozen phonon models across the full scattering-angle range, most prominently between the CAP and the MD-FPM. For diamond, absolute inter-model differences at individual Bragg reflections reach values around $10^{-3}I_0$, with deviations above the statistical convergence threshold persisting up to the HOLZ region near $\theta\approx110\,\text{mrad}$. For STO, the disagreement is substantially more severe, attaining values of the order of units of percent at low scattering angles and remaining significant in the high-angle regime. The enhanced discrepancy in STO is attributed to two compounding effects: the strong atomic-number dependence of TDS, which amplifies the intrinsic mismatch between the CAP and FPM absorption physics for heavy elements; and the anisotropic atomic vibrations of the STO, which neither the CAP nor the isotropic Einstein FPM reproduces.

Nevertheless, we note that the high scattering angle differences might have been affected by the convergence thresholds, as some of the values are below the standard deviation calculated for MD-FPM and in some cases even below the threshold determined by Einstein-FPM convergence.

The relative error at HOLZ and high-angle reflections is disproportionately large in both materials due to the weak absolute intensities at these positions. 
The thickness-dependent analysis further shows that these discrepancies are not static: for STO, the differences between the correlated FPM and both simpler models grow progressively with specimen thickness at high scattering angles, indicating that the approximations underlying the CAP and Einstein FPM accumulate error with increasing propagation depth.

Agreement among all three models can be substantially improved by adopting a large detector collection area. Convolution with a $110\,\text{mrad}$ circular aperture reduces the maximum inter-model discrepancy for diamond to around $10^{-4.5}I_0$, providing quantitative confirmation of the large-detector convergence predicted by Martin et al. and establishing a practical condition under which the computationally efficient CAP model serves as a reliable substitute for the more demanding FPM-MD approach.

In summary, the correlated atomic motion frozen phonon model represents the most physically complete description of TDS within the multislice framework. The CAP model, while advantageous in requiring neither molecular dynamics nor multiple snapshot propagations, introduces systematic errors that are especially pronounced for heavy-element systems with anisotropic vibrations and at the high scattering angles. The conditions under which the CAP provides an adequate approximation — light-element systems, moderate specimen thickness, and large detector collection angles — have been identified and quantified, and may serve as practical guidelines for the selection of an appropriate absorption model in quantitative TEM simulations.